\documentclass[12pt,noshowpacs,nofootinbib,notitlepage,amsmath,superscriptaddress]{revtex4-1}
\usepackage{setspace}
\usepackage{hyperref}
\usepackage{amsfonts}
\usepackage{appendix}
\usepackage{mathtools}
\usepackage{ulem}
\usepackage{ifthen}
\newboolean{PrintVersion}
\setboolean{PrintVersion}{false}
\usepackage{amssymb,amstext} 
\usepackage{graphicx} 
\usepackage[caption=false]{subfig}
\usepackage{wrapfig}
\usepackage{cleveref} 
\usepackage{enumitem}
\usepackage{xcolor}
\usepackage[export]{adjustbox}
\usepackage{float}
\usepackage{booktabs}
\usepackage[section]{placeins}
\usepackage{siunitx}

\ifthenelse{\boolean{PrintVersion}}{   
\hypersetup{	
    citecolor=black,%
    filecolor=black,%
    linkcolor=black,%
    urlcolor=black}
}{} 

\begin{document}
\title{Radial Stability of Charged Interacting Quark Stars Revisited}
\author{Yann Bissessur}
\email{yann.bissessur@ens-paris-saclay.fr}
\affiliation{Université Paris-Saclay, ENS Paris-Saclay, DER de Physique, 91190, 
Gif-sur-Yvette, France}
\affiliation{Department of Physics and Astronomy, University of Waterloo, Waterloo, Ontario, N2L 3G1, Canada}
\author{Michael Gammon}
\email{gammon.michael10@gmail.com}
\affiliation{Department of Physics and Astronomy, University of Waterloo, Waterloo, Ontario, N2L 3G1, Canada}
\author{Robert B. Mann}
\email{rbmann@uwaterloo.ca}
\affiliation{Department of Physics and Astronomy, University of Waterloo, Waterloo, Ontario, N2L 3G1, Canada}


\begin{abstract}
An error in the charged quark star literature \cite{Goncalves:2020joq} has resulted in erroneous solutions to the radial stability equations for charged interacting quark stars in \cite{zhang:2021}.  In this article we recalculate the fundamental radial eigenfrequencies for charged, interacting quark stars in general relativity and discuss any departure from previous results  \cite{zhang:2021}. Most of the qualitative trends persist, but the magnitudes of the separation between the maximum mass and stability points on a given branch can change significantly. We also note that all branches in \cite{zhang:2021} that contained no stable solutions now have stable solutions in their corrected counterparts. The net effect is that in  all of the studied cases charged quark stars have a greater range of stability than originally  predicted. We   also compare four recent observational constraints 
not included in previous studies, and find that these constraints are satisfied by a subset of the mass-radius relations we obtain.

\end{abstract}
\maketitle
\newpage

\section{Introduction}\label{sec:intro}

Thanks to recent advances in observational astrophysics and strong interaction theory, interest in studying compact stars as a tool for understanding high density matter has been rejuvenated \cite{Miller_2019,salmi2024,Horvath_2023,Abbott_2020,Pasechnik:2016wkt,Holdom:2017gdc}. Recent theoretical advances have challenged the Bodmer-Witten-Terazawa hypothesis \cite{Bodmer:1971we,Witten,Terazawa:1979hq} by showing that up-down quark matter ($ud\mathrm{QM}$) might be the ground state of baryonic matter, and thus more stable than even strange quark matter (SQM) \cite{Holdom:2017gdc}. In addition to this, several objects in tension with simple neutron equations of state coupled to general relativity (like PSR J0030+0451 \cite{Miller_2019}, PSR J0740+6620 \cite{salmi2024}, HESS J1731-347 \cite{Horvath_2023}, and GW190814 \cite{Abbott_2020}) have in particular led to a number of articles on quark stars in recent years \cite{Zhang:2020jmb,zhang:2021,gammon_2024,gammon_2025_chargedquarkstars4degb,gammon2026,banerjee_2021_quark,oikonomou_2023_colourflavour,banerjee_2021_strange,panotopoulos2022,Goncalves:2020joq,ren2020,zhang2021_3,panotopoulos2019,Jasim:2021cft}.

In 2021 Zhang, Gammon, and Mann looked at the structure and stability of quark stars with a unified interacting equation of state and a net charge \cite{zhang:2021}, presenting a host of stable charged compact objects with strong interaction effects accounted for in the equation of state \cite{Zhang:2020jmb}.  However, this paper made use of a stability equation for charged compact objects \cite{Goncalves:2020joq} that contained an error. This error, evidently a typo,  does 
 not affect the mass/radius relation for charged quark stars at all.  However the stability properties of these stars can change significantly from what was reported in \cite{zhang:2021}.

In this article we investigate the stability of charged, interacting quark stars, making use of the corrected stability equations. In section \ref{sec:stability} we briefly review the origin of the error  and discuss how the equations governing stability change as a result. In section \ref{sec:resultsposlam} we reproduce the results from \cite{zhang:2021} with the corrected stability data for a positive strong quark interaction constant $\lambda$. The following section correspondingly covers the negative $\lambda$ results. We also discuss the implications of the changes in the results compared to the original 
study
\cite{zhang:2021}. Section \ref{sec:discussion} summarizes qualitative trends from the preceding two sections, and finally section \ref{sec:conclusion} is used for a summary and concluding remarks. For a more thorough look at the theory governing the structure and stability of charged interacting quark stars we direct the reader to \cite{zhang:2021}.

\section{Stability of Charged Interacting Quark Stars}\label{sec:stability}

We consider a static, spherically symmetric metric ansatz
\begin{equation}
     ds^2 = -e^{2\nu(r)} dt^2 + e^{2\Lambda(r)} dr^2
  + r^2\left(d\theta^2 + \sin^2\theta\, d\phi^2\right),
  \label{metric}
\end{equation}
with the quark star interior modeled as a perfect fluid of density $\rho(r)$ and pressure $P(r)$. 
The unified, strongly interacting equation of state we use to relate these quantities is~\cite{Zhang:2020jmb}
\begin{equation}
P(r)=\frac{1}{3}(\rho(r)-4B_{\mathrm{eff}})+ \frac{4\lambda^2}{9\pi^2}\left(-1+\mathrm{sgn}(\lambda)\sqrt{1+3\pi^2 \frac{(\rho(r)-B_{\mathrm{eff}})}{\lambda^2}}\right)
\label{eos_tot}
\end{equation}
where 
\begin{equation}
\lambda=\frac{\xi_{2a} \Delta^2-\xi_{2b} m_s^2}{\sqrt{\xi_4 a_4}}
\label{lam}
\end{equation}
is the generalized strong interaction coupling constant.
Accordingly, the parameter $a_4$ represents the pQCD corrections from one-gluon exchange, $m_s$ denotes the strange quark mass, the
gap parameter $\Delta$ represents the color-superconducting gap,  and the constant $\xi$ coefficients take on particular sets of values depending on the phase of the quark matter, namely
\begin{align}
(\xi_4,\xi_{2a}, \xi_{2b}) = \left\{ \begin{array} {ll}
(( \left(\frac{1}{3}\right)^{\frac{4}{3}}+ \left(\frac{2}{3}\right)^{\frac{4}{3}})^{-3},1,0) & \textrm{2SC phase}\\
(3,1,3/4) & \textrm{2SC+s phase}\\
(3,3,3/4)&   \textrm{CFL phase}
\end{array}
\right.
\end{align}
Note that   $\lambda > 0$ provided $\Delta^2/m_s^2>\xi_{2b}/\xi_{2a}$. The parameter $B_{\rm eff}$ is the effective bag constant that accounts for the nonperturbative contribution from the QCD vacuum.  In the present manuscript we assume a typical value  $B_\mathrm{eff}=60 \; \si{MeV/fm^3}$ in order to be consistent with \cite{zhang:2021}. However, our results are also presented in a unitless form which can in principle be rescaled with any desired value of $B_\mathrm{eff}$.

The Tolman-Oppenheimer-Volkov (TOV) equations~\cite{Oppenheimer:1939ne,Tolman:1939jz} with charge effects included~\cite{Ray:2003gt} take the form:
\begin{align}
  \frac{\mathrm{d}q}{\mathrm{d}r} & {} = 4\pi r^2 \rho_e e^{\Lambda} \, \label{eq:TOV1} \\
  \frac{\mathrm{d}m}{\mathrm{d}r} & {} = 4\pi r^2 \rho  + \frac{q}{r}\frac{\mathrm{d}q}{\mathrm{d}r} \,  \label{eq:TOV2} \\
  \frac{\mathrm{d}P}{\mathrm{d}r} & {} = -(\rho + P)\left(4\pi r P + \frac{m}{r^2} - \frac{q^2}{r^3}\right)e^{2\Lambda}  
  + \frac{q}{4\pi r^4}\frac{\mathrm{d}q}{\mathrm{d}r}   \label{eq:TOV3} \\
   \frac{\mathrm{d}\nu}{\mathrm{d}r}  & {} = -\frac{1}{\rho + P}\left(\frac{\mathrm{d}P}{\mathrm{d}r} - \frac{q}{4\pi r^4}\frac{\mathrm{d}q}{\mathrm{d}r}\right) \, \label{eq:TOV4}
  \end{align}
where $q(r)$ and $m(r)$,  respectively, represent the charge and mass within radius $r$, $\rho_e(r)$ is the electric charge density at $r$, and
\begin{equation}
  \label{chargedTOV}
  e^{-2\Lambda(r)} = 1 - \frac{2m(r)}{r} + \frac{q(r)^2}{r^2}\,
\end{equation}
is the radial metric function.

The stability of gravitationally bound compact objects is usually studied by examining their fundamental mode eigenfrequencies. This is done by assuming a fluid element is perturbatively  displaced from its equilibrium position $r$ to $r + \delta r$, and that the perturbation has a harmonic time dependence (namely $e^{i\omega t}$). Under spherical symmetry the equations describing the perturbation are of the Sturm-Liouville form \cite{Brillante:2014lwa}, namely:
\begin{equation}\label{SLform}
\frac{d}{d r}\left[\mathcal{P} \frac{d u}{d r}\right]+\left[\mathcal{Q}+\omega^2 \mathcal{W}\right] u=0,
\end{equation}
or equivalently \cite{zhang:2021}:
\begin{equation}
    \begin{aligned}
        \frac{d u}{dr}&=\frac{\eta}{\mathcal{P}}, \\
        \frac{d \eta}{d r}&=-\left[\mathcal{Q}+\omega^2 \mathcal{W}\right] u,
    \end{aligned}
\end{equation}
where $u(r)=r^2e^{-\nu(r)} \delta r$ is the normalized displacement function. 

The functions $\mathcal{P}$, $\mathcal{Q}$, $\mathcal{W}$ are 
\cite{Brillante:2014lwa} 
\begin{equation}
\begin{aligned}
\mathcal{P} & =e^{\Lambda+3 \nu} r^{-2} \gamma P \\
\mathcal{Q} & =\left(\rho+P\right)\Big( r^{-2} \left(\nu^\prime\right)^2 e^{\Lambda+3 \nu}-8 \pi r^{-2} e^{3 \nu+3 \Lambda} P -r^{-6} e^{3 \nu+3 \Lambda} q^2\Big) -4 r^{-3} P^\prime e^{\Lambda+3 \nu} \\
\mathcal{W} & =e^{3 \Lambda+\nu} r^{-2}\left(\rho+P\right)
\end{aligned}
\end{equation}
where $\gamma=(1+\frac{\rho}{P})\frac{dP}{d\rho}$ is the adiabatic index.  
Making use of equation  \eqref{eq:TOV3} 
we arrive at a simplified final form for the $\mathcal{Q}$ equation: 
\begin{align}\label{Qcorr}
    \mathcal{Q} &= (\rho + P)r^{-2}e^{\Lambda + 3\nu}\Bigg[\nu^\prime\left(\nu^\prime + 4r^{-1}\right)- (8\pi P  + r^{-4}q^2)e^{2\Lambda}\Bigg] - \frac{q q^\prime}{\pi r^7} e^{\Lambda+3\nu}.
\end{align}

This is notably different than equation (14) in \cite{Goncalves:2020joq} which reads:
\begin{equation}\label{eq:typo}
    \mathcal{Q}=(\rho+P) r^{-2} e^{\Lambda+3 \nu}\left[\nu^\prime\left(\nu^\prime-4 r^{-1}\right)-\left(8 \pi P+r^{-4} q^2\right) e^{2 \Lambda}\right].
\end{equation}
Written this way, $\mathcal{Q}$ is missing the $-\frac{q q'}{ \pi r^7}e^{\Lambda+3\nu}$ term and has a flipped sign on the term linear in $\nu'$. The  earlier stability analysis \cite{zhang:2021}
employed \eqref{eq:typo}
but with the correct (plus) sign for the $\nu'$ term.

From here, as before, radial stability of these stars can be determined by the sign of the fundamental mode eigenfrequency $\omega_0^2$ - radial stability is achieved when  $\omega_0^2\geq 0$. For uncharged compact stars, the point at which the fundamental radial mode eigenfrequency switches signs is always coincident with the point where $\partial M/\partial \rho_\mathrm{c}$ changes sign (here $\rho_\mathrm{c}$ is the central mass density of a star, a boundary condition). The result of this is the zero fundamental eigenfrequency point coinciding with the curve's maximum mass point.

For charged compact stars this coincidence is no longer guaranteed. The specifics of this change (whether the stability transition point moves to higher/lower central densities, and by how much) depend on the model chosen for the distribution of charge, and the strength of the strong interaction parameter $\lambda$. We use the following section to share these specifics.

 For reference we have also included four recent observational constraints in our $M$/$R$ diagrams (not present in \cite{zhang:2021}), corresponding to the interesting astrophysical observations discussed in section \ref{sec:intro}. These are objects that are in some tension with common neutron star EOSs coupled to general relativity, and thus they are interesting in the context of exotic compact objects. More specifically, in \cref{fig:modelAresults,fig:modelBresults,fig:modelCresults,fig:mA_negLam,fig:mB_negLam,fig:mC_negLam} the coloured boxes are $1\sigma$ observational estimates of mass/radii for PSR J0030+0451 (violet) \cite{Miller_2019}, PSR J0740+6620 (yellow) \cite{salmi2024}, HESS J1731-347 (gray) \cite{Horvath_2023}, and GW190814 (green) \cite{Abbott_2020}. It is intriguing that a subset of the observed mass-radius constraints for all four objects are consistent with the curves we obtain for all three charge models.

\section{Results for Positive $\lambda$}\label{sec:resultsposlam}
Here we present the results describing the stellar structure and the radial stability for charged, interacting quark stars with a positive strong interaction constant $\lambda$.
 In all cases the mass/radius results are unchanged (matching those in \cite{zhang:2021});  because of this we primarily focus on the location of the `stability point' (the point at which $\omega_0^2$  changes signs) relative to the maximum mass point along the mass-radius curve.
 
 \subsection{Charge Model A}\label{sec:modelAplus}
\begin{figure}[h]
 \centering
 \includegraphics[width=8 cm]{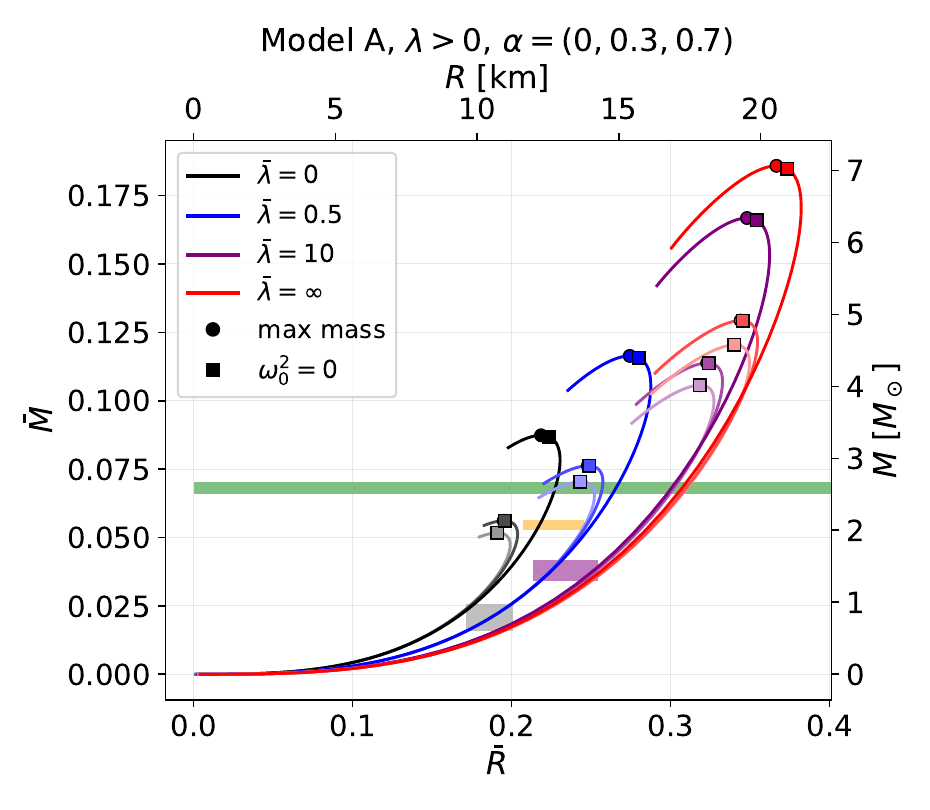}  
\includegraphics[width=8 cm]{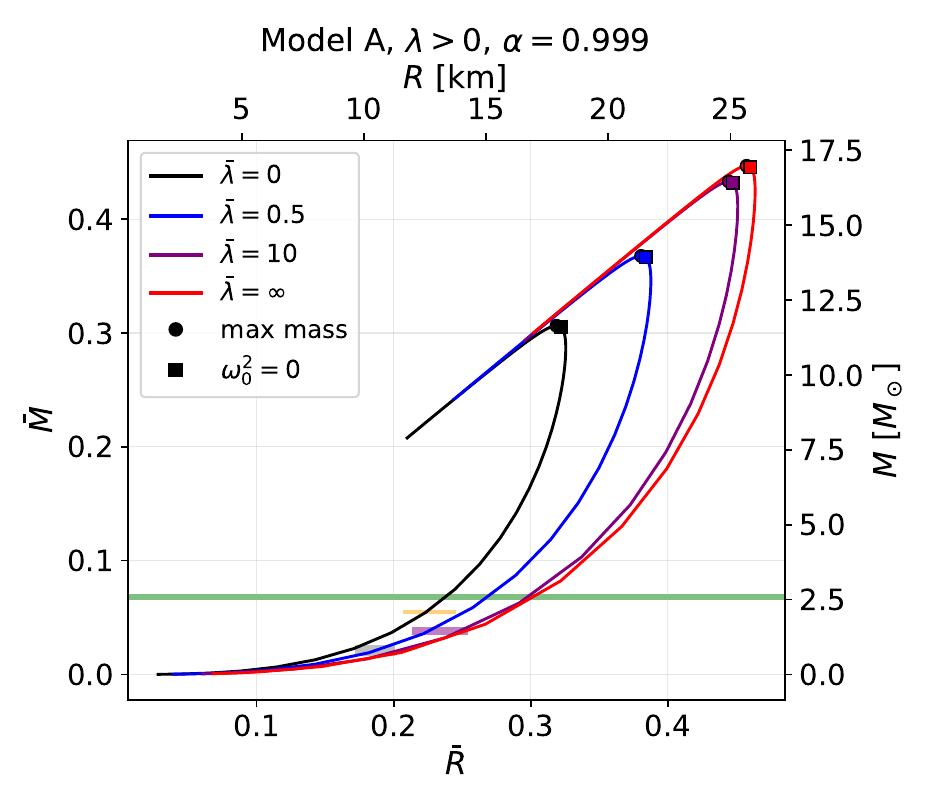}  
   \includegraphics[width=8 cm]{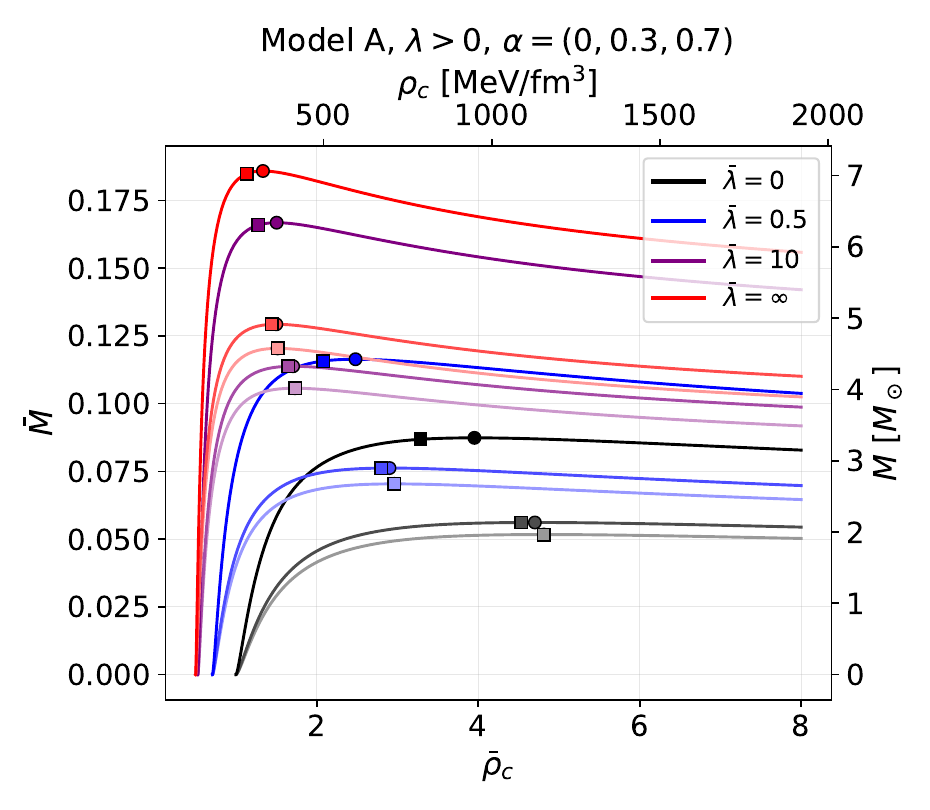}  
  \includegraphics[width=8 cm]{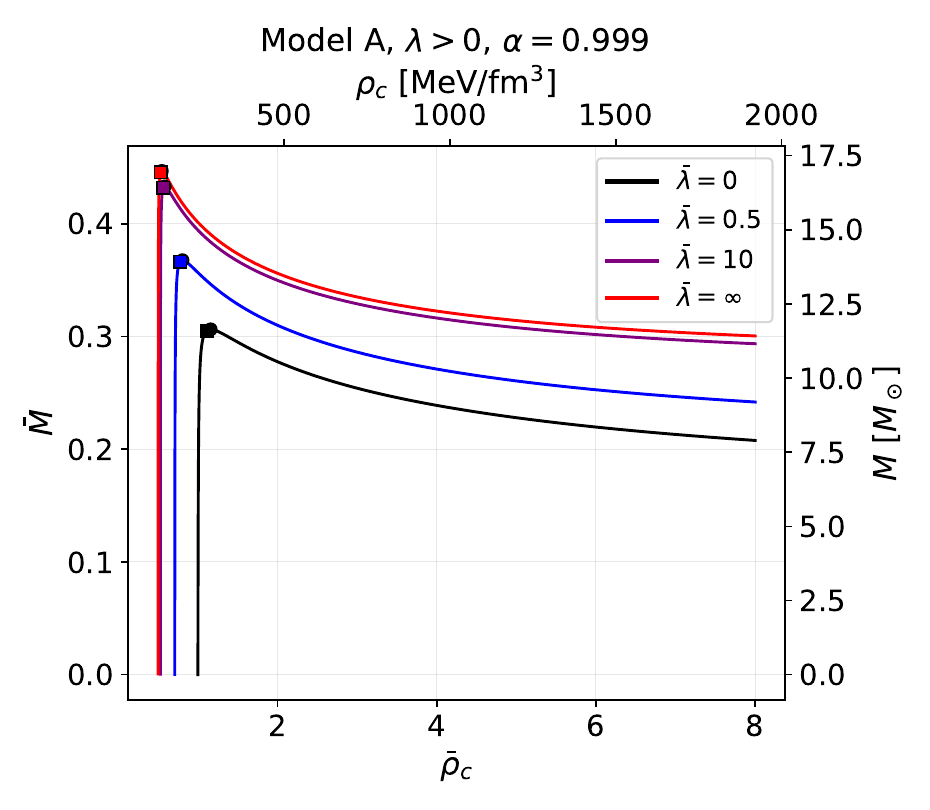}  
\caption{Plots of $\bar{M}$ vs. $\bar{R}$ (upper graphs) and $\bar{M}$ vs. $\bar{\rho}_c$ (lower graphs) of charged interacting quark stars in model A for (left) $\alpha=(0, 0.3, 0.7)$ and (right) $\alpha=0.999$. The right and top axes in each plot are the corresponding dimensional results with $B_{\rm eff}=60\, \si{MeV/fm^3}$ for illustration and consistency with \cite{zhang:2021}.
The  black, blue, violet, and red curves respectively denote   $\bar{\lambda}=(0, 0.5 , 10, \infty)$ with the sign of $\lambda$ being positive. Darker shades correspond to increasing values of $\alpha$.
The solid dots denote the maximum mass configurations, with the filled squares denoting the point at which $\bar{\omega}^2_0=0$.  Finally, the coloured boxes are $1\sigma$ observational estimates of mass/radii for PSR J0030+0451 (violet) \cite{Miller_2019}, PSR J0740+6620 (yellow) \cite{salmi2024}, HESS J1731-347 (gray) \cite{Horvath_2023}, and GW190814 (green) \cite{Abbott_2020}.} 

\label{fig:modelAresults}
\end{figure}

 Charge model A assumes a proportionality
 $$
 \rho_e=\alpha\rho
 $$
 between mass density and charge density \cite{zhang:2021,Goncalves:2020joq,Ray:2003gt,panotopoulos2019} (or, in unitless form, $\bar{\rho}_\mathrm{e}=\alpha \bar{\rho}$ - see appendix \ref{sec:appendix}). These results are shown along with the associated $M$/$R$ and $M/\rho_\mathrm{c}$ curves in figure \ref{fig:modelAresults}.

 As before, the stability point coincides with the maximum mass point for the uncharged solution, and as before when charge is introduced the stability point separates from the maximum mass point and moves to a star with smaller central density. 
We find, however, that the relative size of the gap  is much smaller than previously reported
\cite{zhang:2021}. This means that more of these stars are stable than we previously thought. The effect of increasing $\lambda$ is a suppression of the separation induced by the charge, both in \cite{zhang:2021} and in figure \ref{fig:modelAresults}. 
 
Furthermore, contrary to the previous study \cite{zhang:2021}, we also see a turnaround behaviour in the stability point/maximum mass separation. By the time we get to very large charge ($\alpha=0.999$) the stability point has returned to coincidence with the maximum mass point, regardless of $\lambda$ - this is qualitatively different behaviour than what was predicted in \cite{zhang:2021}, where the separation continued to grow with charge all the way to $\alpha=0.999$. Once again we note that there are more stable stars in the parameter space than was previously thought.

 \subsection{Charge Model B}
 
\begin{figure}[h]
 \centering
 \includegraphics[width=8 cm]{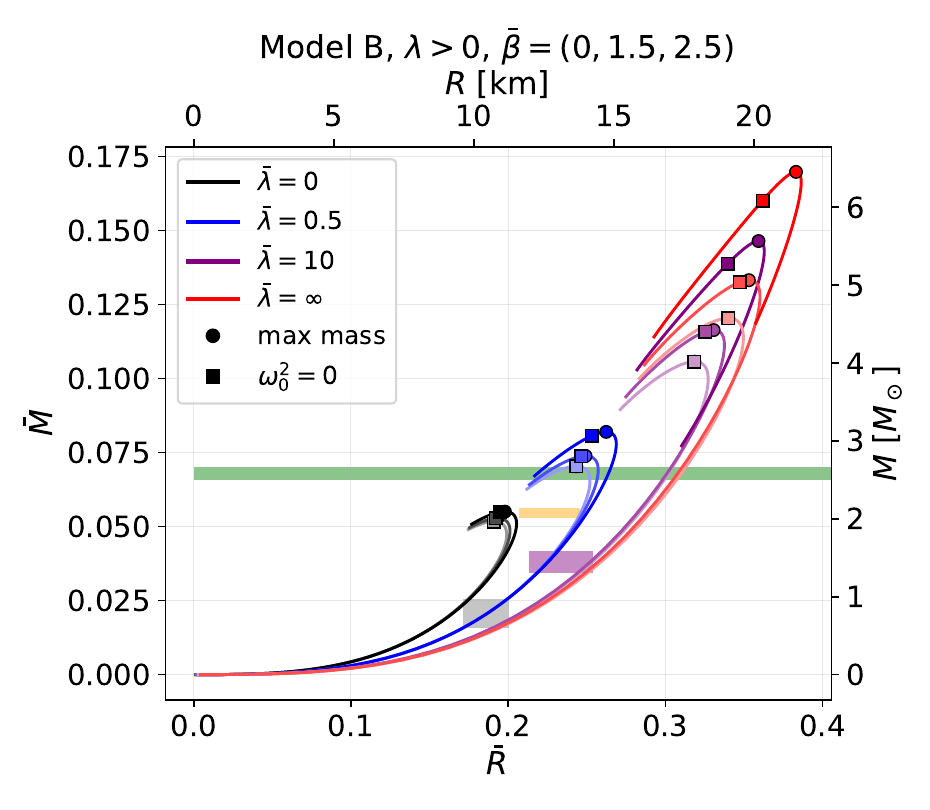}  
  \includegraphics[width=8 cm]{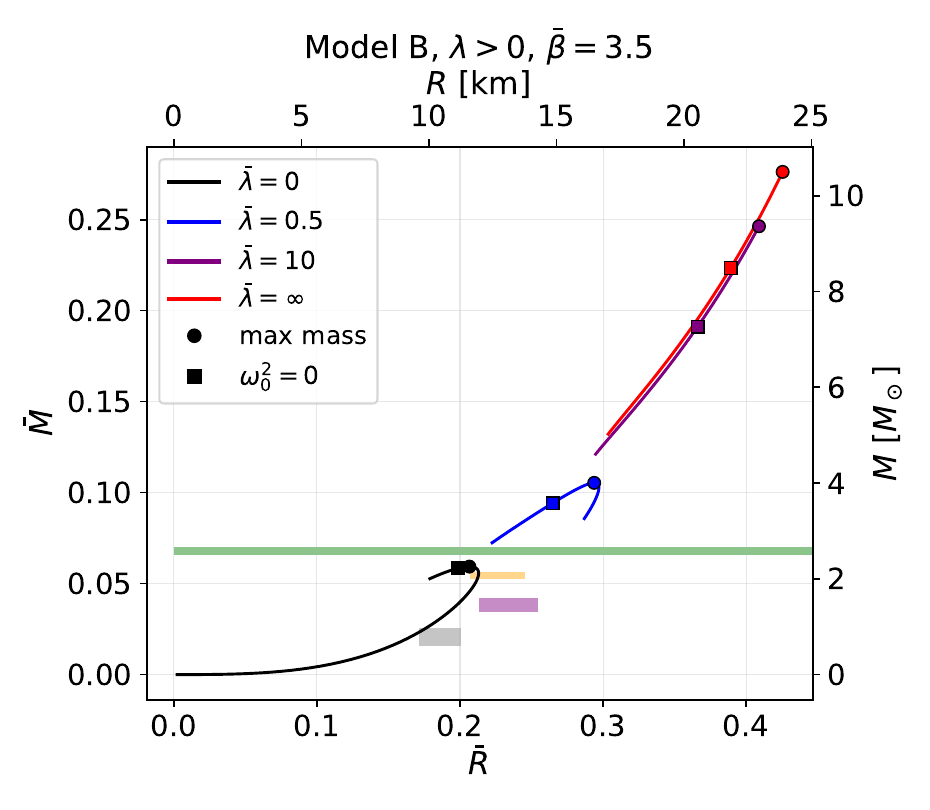}  
   \includegraphics[width=8 cm]{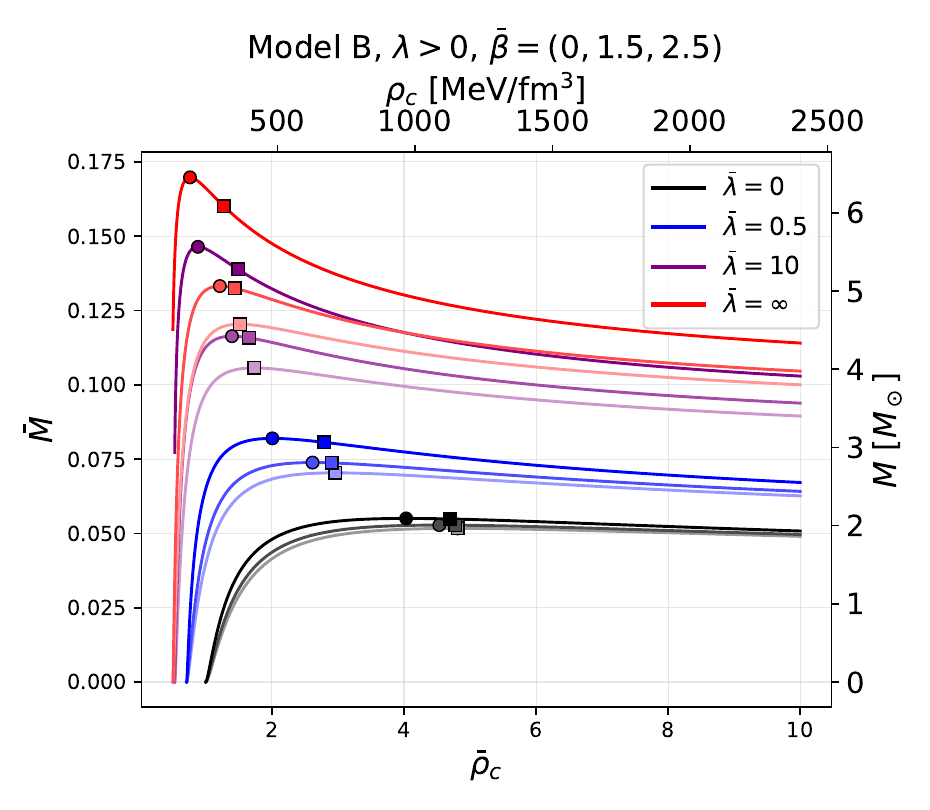}  
  \includegraphics[width=8 cm]{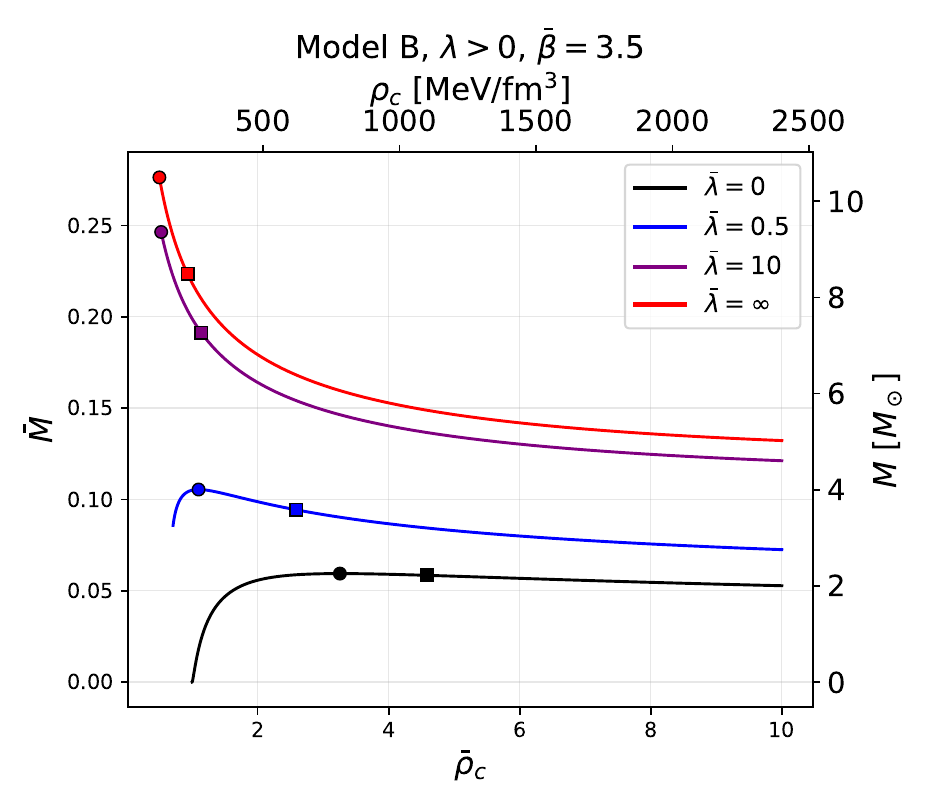}  
\caption{Plots of $\bar{M}$ vs. $\bar{R}$ (upper graphs) and $\bar{M}$ vs. $\bar{\rho}_c$ (lower graphs) of charged interacting quark stars in model B for (left) $\bar{\beta}=(0,1.5, 2.5)$   and (right) $\bar{\beta}=3.5$. The right and top axes in each plot are the corresponding dimensional results with $B_{\rm eff}=60\, \si{MeV/fm^3}$ for illustration and consistency with \cite{zhang:2021}.  The  black, blue, violet, and  red curves respectively denote  $\bar{\lambda}=(0, 0.5 , 10, \infty)$ with the sign of $\lambda$ being positive. Darker shades correspond to increasing values of $\bar{\beta}$.
The solid dots denote the maximum mass configurations, with the filled squares denoting the point at which $\bar{\omega}^2_0=0$. Finally, the coloured boxes are $1\sigma$ observational estimates of mass/radii for PSR J0030+0451 (violet) \cite{Miller_2019}, PSR J0740+6620 (yellow) \cite{salmi2024}, HESS J1731-347 (gray) \cite{Horvath_2023}, and GW190814 (green) \cite{Abbott_2020}.
}
   \label{fig:modelBresults}
\end{figure}

Charge model B directly assumes that total charge is proportional to spatial volume \cite{zhang:2021,Goncalves:2020joq,Arbanil:2015uoa} via 
$$
q(r)=Q\left(\frac{r}{R}\right)^3\equiv\beta r^3
$$ 
(or in unitless form $\bar{q}(\bar{r})=\bar{\beta}\bar{r}^3$ - see appendix \ref{sec:appendix}). The $M$/$R$ and $M$/$\rho_\mathrm{c}$ results for this charge model are shown in figure \ref{fig:modelBresults}. 

Commensurate with the previous study \cite{zhang:2021}, charge model B offsets the stability point in the opposite direction relative to charge model A (namely to larger central densities) as a nonzero charge parameter is introduced. Here this separation continues growing as $\beta$ increases - we do not observe the turnaround behaviour that was present for model A, meaning that the largest separation between maximum mass and the stability point is seen at the largest $\beta$ considered. 

Furthermore, the previous study
\cite{zhang:2021} reported that the  size of the central density separation  gets rapidly larger, then slowly becomes smaller as $\bar{\lambda}$ increases. We also observe similar behaviour here - though difficult to tell by eye,   in figure \ref{fig:modelBresults} the central density separation also follows this trend (namely growing from $\bar{\lambda}=0\to\bar{\lambda}=0.5$, and then shrinking for larger $\lambda$). In general we also see a larger average separation size,   which means the sequence contains more stable stars than reported in \cite{zhang:2021}. We also note the  truncations of the curves in the upper right panel, which  appear because of the exotic pressure profiles associated with charge model B (ie. the pressure first increases above its central value $p_\mathrm{c}$ before decreasing to 0 at finite radius). The net effect is that the $M$/$R$ curves   do not start from the origin. This behaviour was also noted in the previous study \cite{zhang:2021} where a more thorough discussion   is included.

Finally,  the curves for largest $\beta$   were originally found to have   no stability point for large enough $\bar{\lambda}$  \cite{zhang:2021} -- it was reported that the fundamental mode eigenfrequency was negative for all solutions along such curves. As we see in figure \ref{fig:modelBresults} this is no longer the case  -- all branches have at least some stable solutions, including those branches that don't start from the origin.

\subsection{Charge Model C}

\begin{figure}[h]
 \centering
 \includegraphics[width=8 cm]{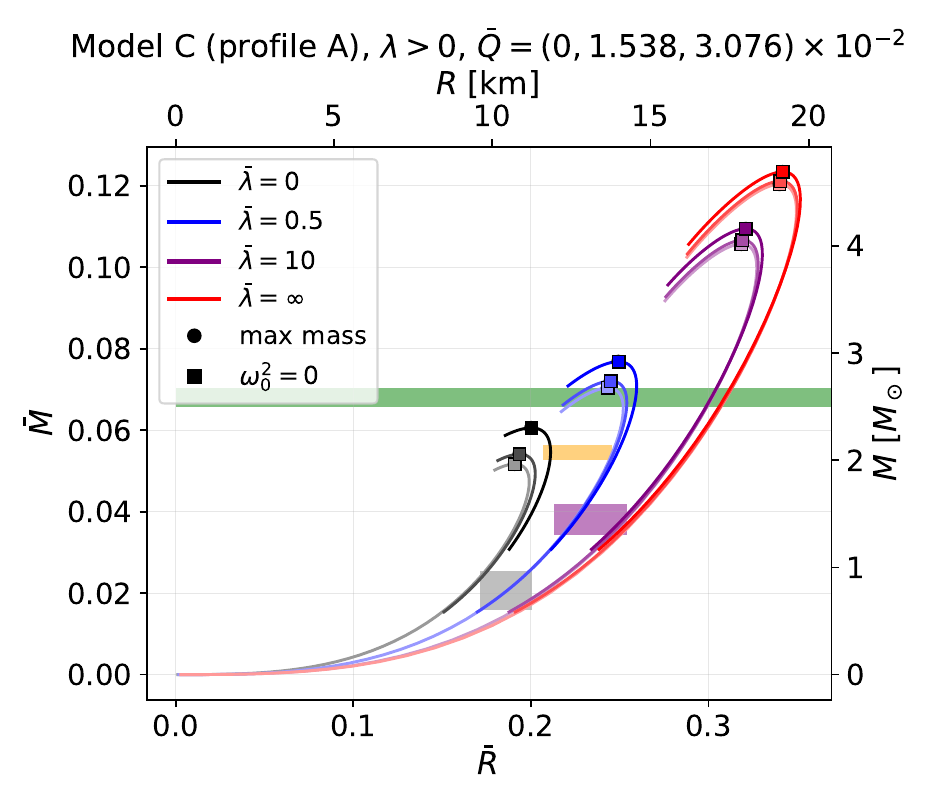}  
  \includegraphics[width=8 cm]{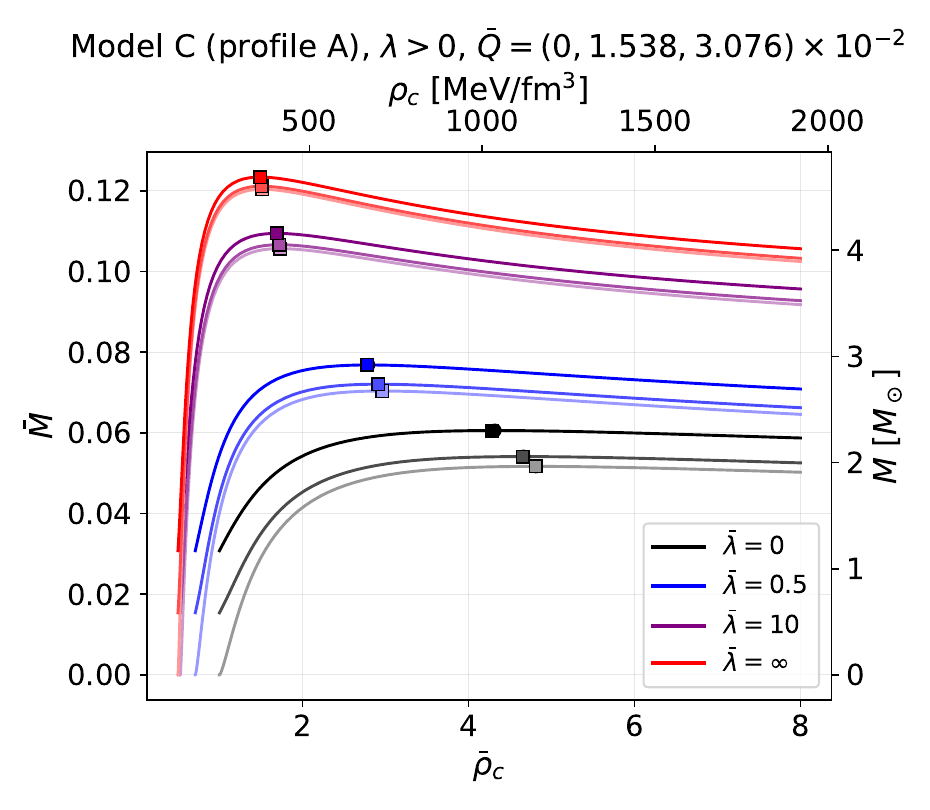}  
\caption[Short caption]{ $\bar{M}$ vs. $\bar{R}$ (left) and $\bar{M}$ vs. $\bar{\rho}_c$ (right) plots of charged interacting quark stars in model C for $\bar{Q}=(0, 1.538,3.076)\times 10^{-2}$. The right and top axes in each plot are the corresponding dimensional results with $B_{\rm eff}=60\, \si{MeV/fm^3}$ for illustration and consistency with \cite{zhang:2021}. The  black, blue, violet, and red curves respectively denote  $\bar{\lambda}=(0, 0.5 , 10,\infty)$ with the sign of $\lambda$ being positive. Darker shades correspond to increasing values of $\bar{Q}$.
The solid dots denote the maximum mass configurations, with filled squares denoting the point at which $\bar{\omega}^2_0=0$. Finally, the coloured boxes are $1\sigma$ observational estimates of mass/radii for PSR J0030+0451 (violet) \cite{Miller_2019}, PSR J0740+6620 (yellow) \cite{salmi2024}, HESS J1731-347 (gray) \cite{Horvath_2023}, and GW190814 (green) \cite{Abbott_2020}.
}
\label{fig:modelCresults}
\end{figure}

In this case we fix the total charge, whose actual values ($\bar{Q}=(0, 1.538,3.076)\times 10^{-2}$, corresponding to $Q=(0,1, 2)\times 10^{20} \; \si{C}$ for a typical bag constant value $B_{\rm eff}=\rm 60\, MeV/fm^3$) are chosen to be commensurate with past work \cite{zhang:2021,Goncalves:2020joq}. The corresponding results are obtained by enumerating the charge configurations and central densities that yield the chosen fixed charge value for a given $\bar{\lambda}$. This can be implemented with interior charge distributions of type model A or model B; we choose to use the former model (in line with \cite{zhang:2021}) for the results in figure \ref{fig:modelCresults}\footnote{We find for the same fixed charges, that the results for $(\bar{M},\bar{R})$ derived from model A differ from those derived from model B on the order of $0.1\%$-$1\%$, with the former having slightly larger maximum masses.}, which is consistent with the stability points shifting to lower central densities rather than higher.  As in model B, some of the curves truncate at smaller mass/central pressure. This is due to charge model C fixing the star's total charge rather than the charge model parameter $\alpha / \beta$. Because of this, an arbitrarily small star can't necessarily support a large predetermined charge. The  curves are thus `cut off' when no solutions have the required total charge.

Here the separations are much smaller than the equivalent plot in \cite{zhang:2021} (which aligns with the results in \ref{sec:modelAplus}), which indicates that there are more stable stars  than previously reported. In the present work the fixed charges are too small to appreciably offset the stability point from the maximum mass point - the differences we do see are on the order of $<0.1\%$ for central densities, $<0.01\%$ for the radii, and $<10^{-7}$ for the unitless masses. 

\section{Results for Negative $\lambda$}\label{sec:resultsneglam}
\begin{figure}[htb]
 \centering
  \includegraphics[width=8 cm]{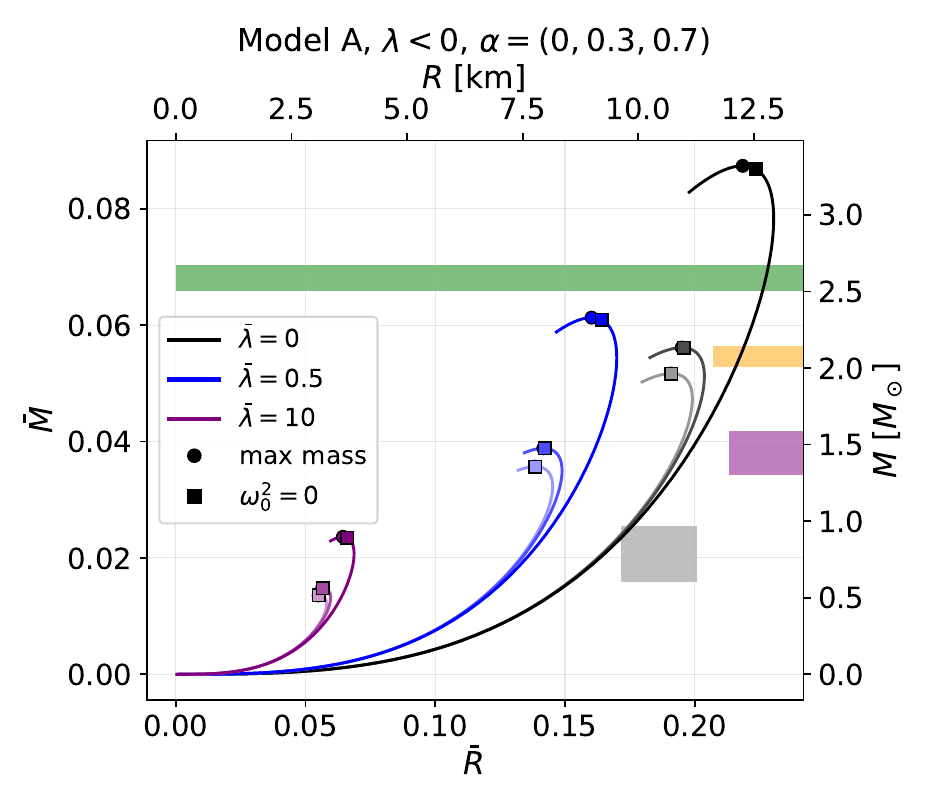}  
    \includegraphics[width=8 cm]{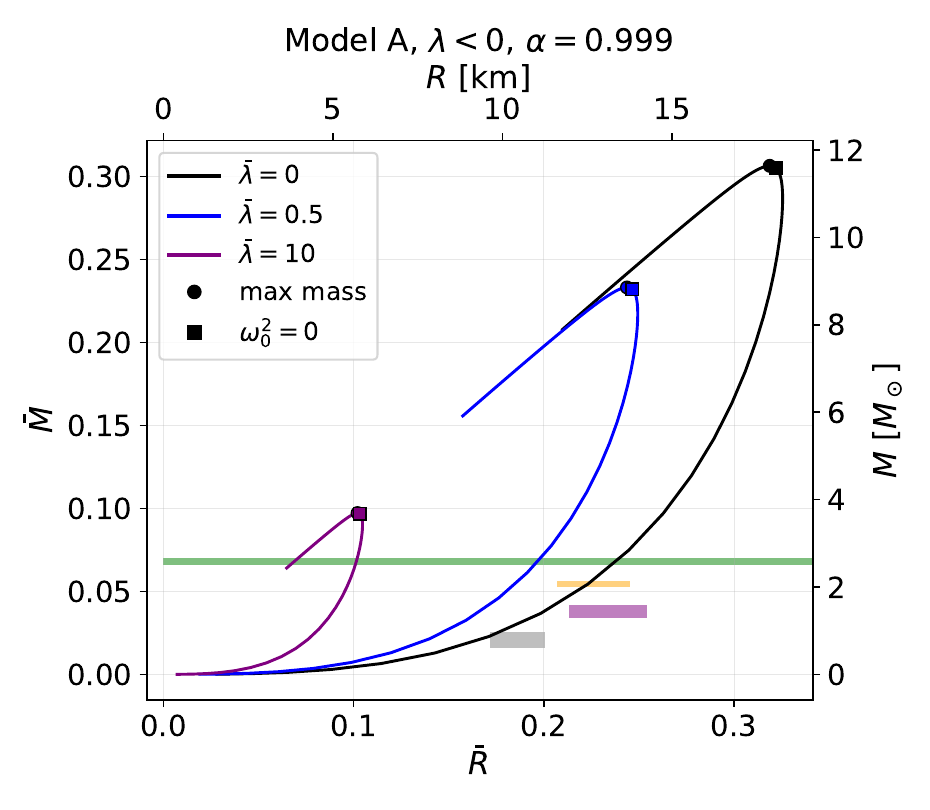}  
   \includegraphics[width=8 cm]{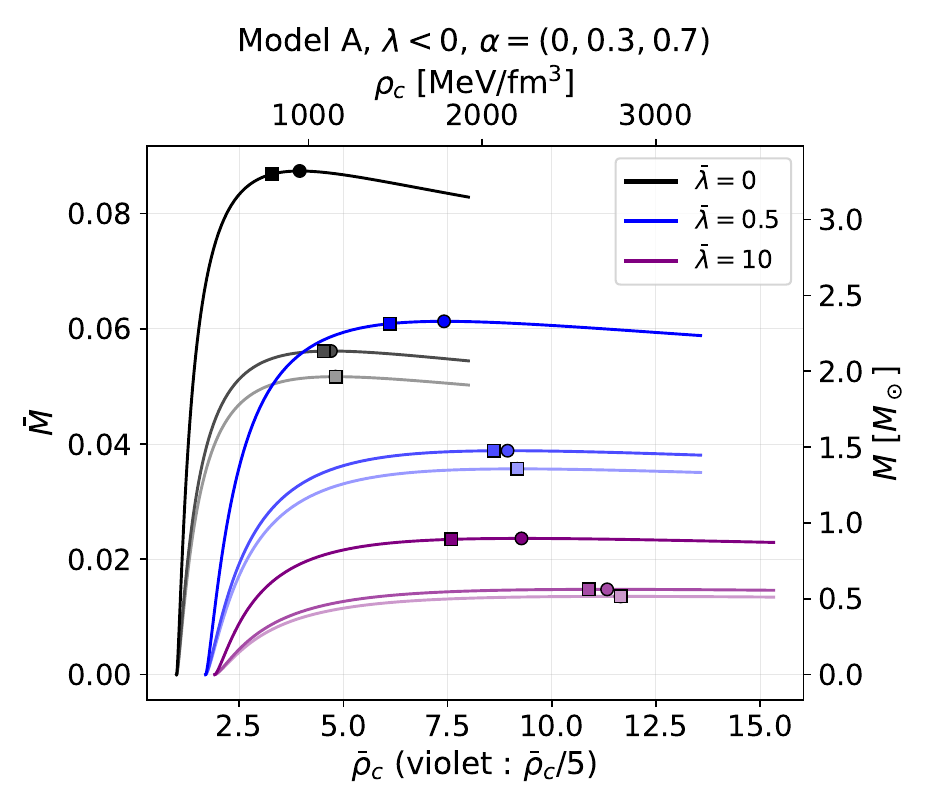}  
      \includegraphics[width=8 cm]{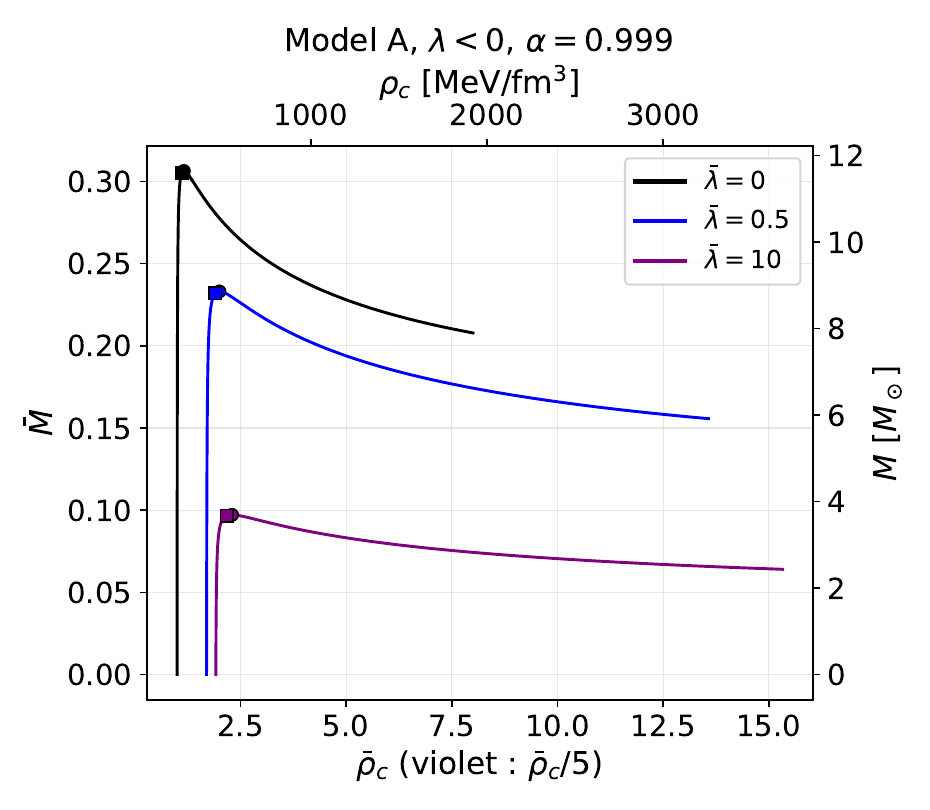}  
\caption{Plots of $\bar{M}$ vs. $\bar{R}$ (upper graphs) and $\bar{M}$ vs. $\bar{\rho}_c$ (lower graphs) of charged interacting quark stars in model A for (left) $\alpha=(0, 0.3, 0.7)$ and (right) $\alpha=0.999$. The right and top axes in each plot are the corresponding dimensional results with $B_{\rm eff}=60\, \si{MeV/fm^3}$ for illustration and consistency with \cite{zhang:2021}. The  black, blue, and violet curves respectively denote $\bar{\lambda}=(0, 0.5 , 10)$ with the sign of $\lambda$ being negative. Darker shades correspond to increasing values of $\alpha$. The solid dots denote the maximum mass configurations, with the filled squares representing where $\bar{\omega}^2_0=0$.  Finally, the coloured boxes are $1\sigma$ observational estimates of mass/radii for PSR J0030+0451 (violet) \cite{Miller_2019}, PSR J0740+6620 (yellow) \cite{salmi2024}, HESS J1731-347 (gray) \cite{Horvath_2023}, and GW190814 (green) \cite{Abbott_2020}. Note that for the violet curves in the $\bar{M}$ vs. $\bar{\rho}_c$ plots, we rescaled the x-axis as $\bar{\rho}_c\to\bar{\rho}_c/5$ for clarity.}
\label{fig:mA_negLam}
\end{figure}

In the previous treatment of model A with $\lambda<0$ \cite{zhang:2021}, the added charge was found to always induce a separation between the two points of interest, and as the magnitude of  (negative) $\lambda$ grew, an increase was seen in the separation between points in central density space. The same behaviour is seen here, but with much smaller density separations on average than those presented in \cite{zhang:2021} (see figure \ref{fig:mA_negLam}). As expected in model A,  the maximum mass points are  at larger central densities than the stability points (in both \cite{zhang:2021} and the present work). This density separation concurrently manifests as a decrease in separation in $M$/$R$ space in \cite{zhang:2021}, which we find is also the case here, just on smaller separation scales. This isn't inconsistent with the former result as a large change in central density doesn't necessarily correspond to a large change in mass.
  However, unlike \cite{zhang:2021}, we once again see that the effect of large charge in model A is to force the separation back to 0, 

even when $\lambda$ is negative and has a large magnitude. In the previous study \cite{zhang:2021} the large charge limit was found to have very low differences in central density between maximum mass and stability points, but had a large mass separation between those same points. Employing the corrected stability equations, we do not observe this.   These new predictions allow for more stable stars that are strongly charged.

\begin{figure}[h]
 \centering
 \includegraphics[width=8 cm]{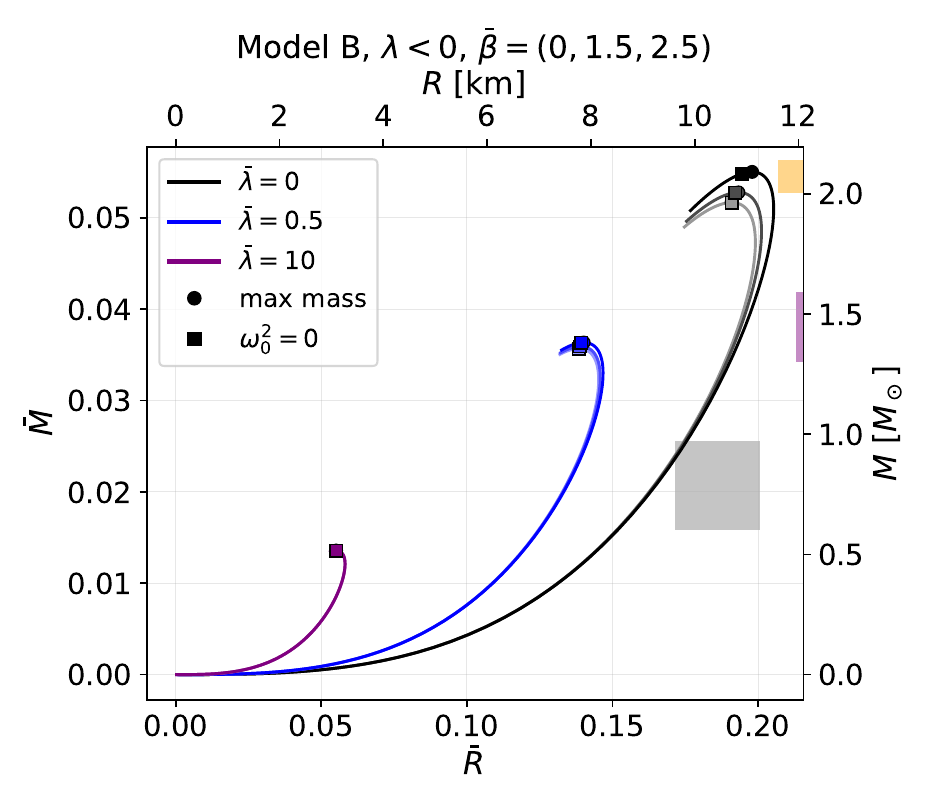}  
     \includegraphics[width=8 cm]{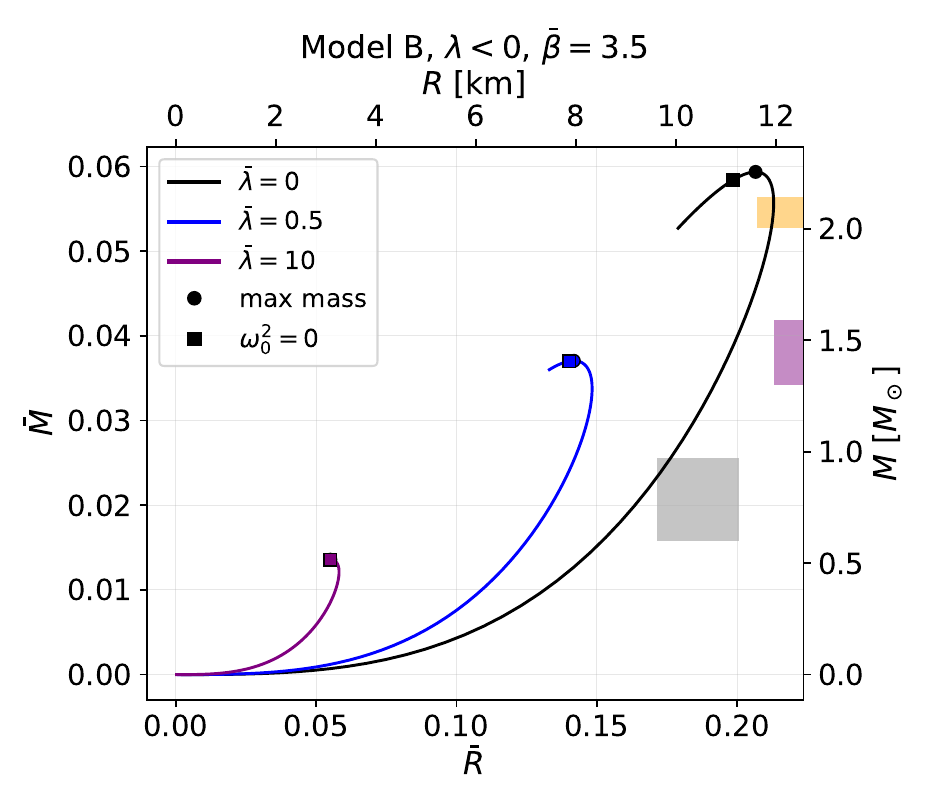}  
   \includegraphics[width=8 cm]{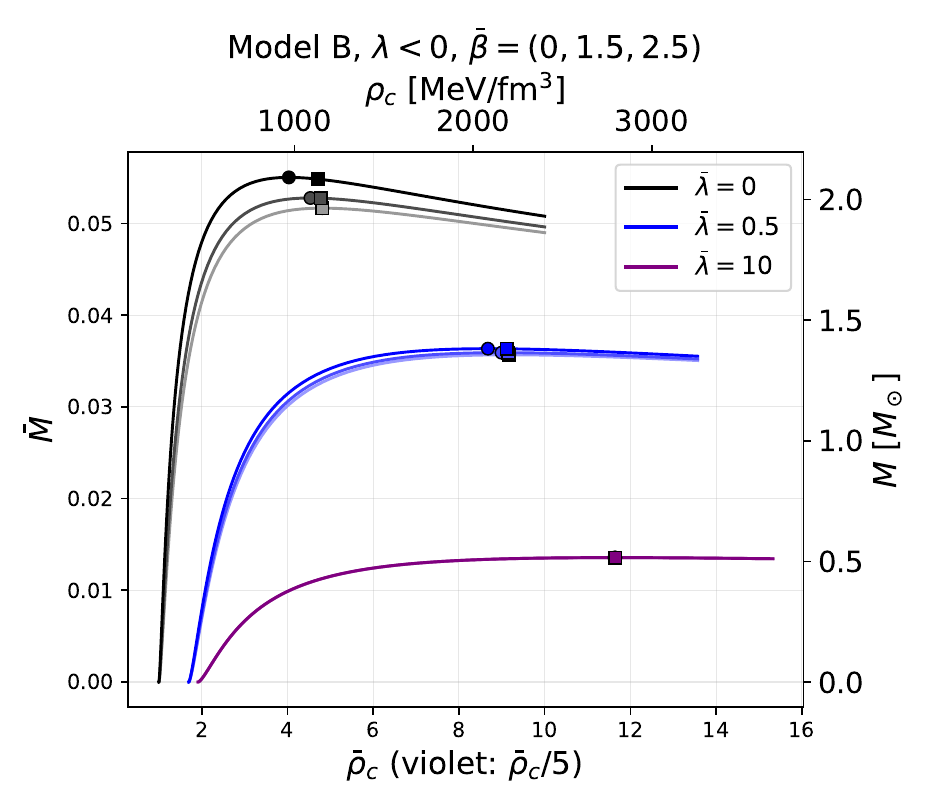}  
     \includegraphics[width=8 cm]{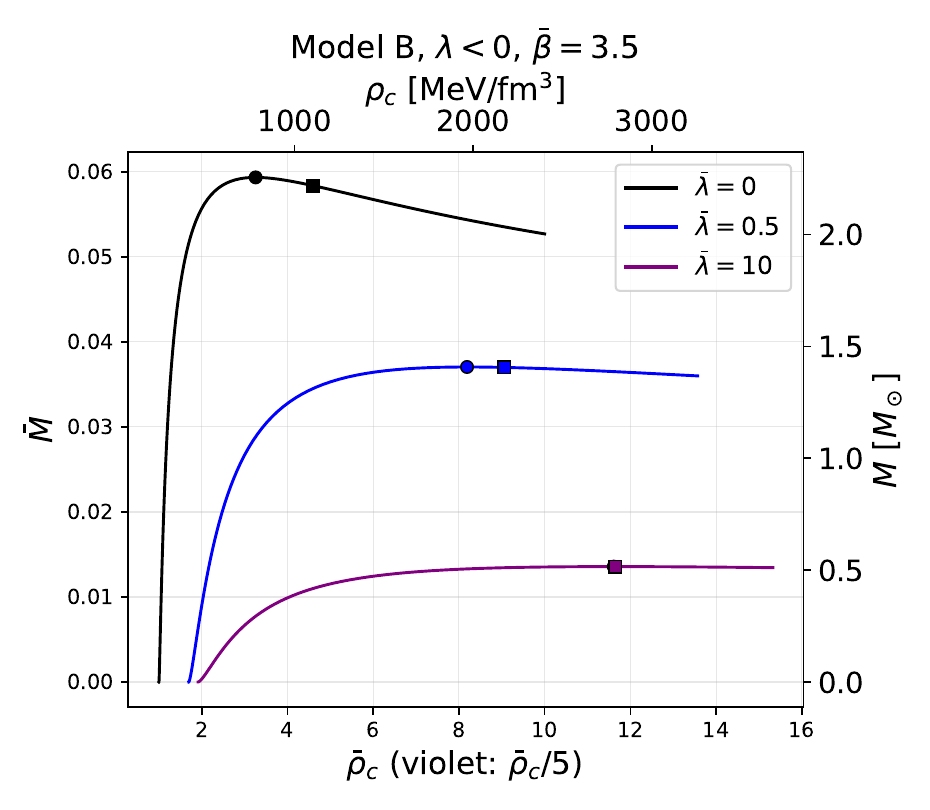}  
\caption{Plots of $\bar{M}$ vs. $\bar{R}$ (upper graphs) and $\bar{M}$ vs. $\bar{\rho}_c$ (lower graphs) of charged interacting quark stars in model B for (left) $\bar{\beta}=(0,1.5, 2.5)$  and (right) $\bar{\beta}=3.5$. The right and top axes in each plot are the corresponding dimensional results with $B_{\rm eff}=60\, \si{MeV/fm^3}$ for illustration and consistency with \cite{zhang:2021}. The black, blue, and violet curves respectively denote $\bar{\lambda}=(0, 0.5 , 10)$ with the sign of $\lambda$ being negative. Darker shades correspond to increasing values of $\bar{\beta}$. 
The solid dots denote the maximum mass configurations, with the filled squares representing where $\bar{\omega}^2_0=0$.  Finally, the coloured boxes are $1\sigma$ observational estimates of mass/radii for PSR J0030+0451 (violet) \cite{Miller_2019}, PSR J0740+6620 (yellow) \cite{salmi2024}, HESS J1731-347 (gray) \cite{Horvath_2023}, and GW190814 (green) \cite{Abbott_2020}. Note that for the violet curves in the $\bar{M}$ vs. $\bar{\rho}_c$ plots, we rescaled the x-axis as $\bar{\rho}_c\to\bar{\rho}_c/5$ for clarity.}
\label{fig:mB_negLam}
\end{figure}

For model B with $\lambda<0$, introducing charge was found to offset the stability point to higher central densities \cite{zhang:2021}  (as it does here), but only appreciably so when the magnitude of $\lambda$ is small. This means that making $\lambda$ more negative here suppresses the separation induced by charge model B. We see analogous behaviour in figure \ref{fig:mB_negLam}, but with larger average separations than those in \cite{zhang:2021}. This means that relative to \cite{zhang:2021} we predict more stable stars in the sequence.

\begin{figure}[h]
 \centering
   \includegraphics[width=8 cm]{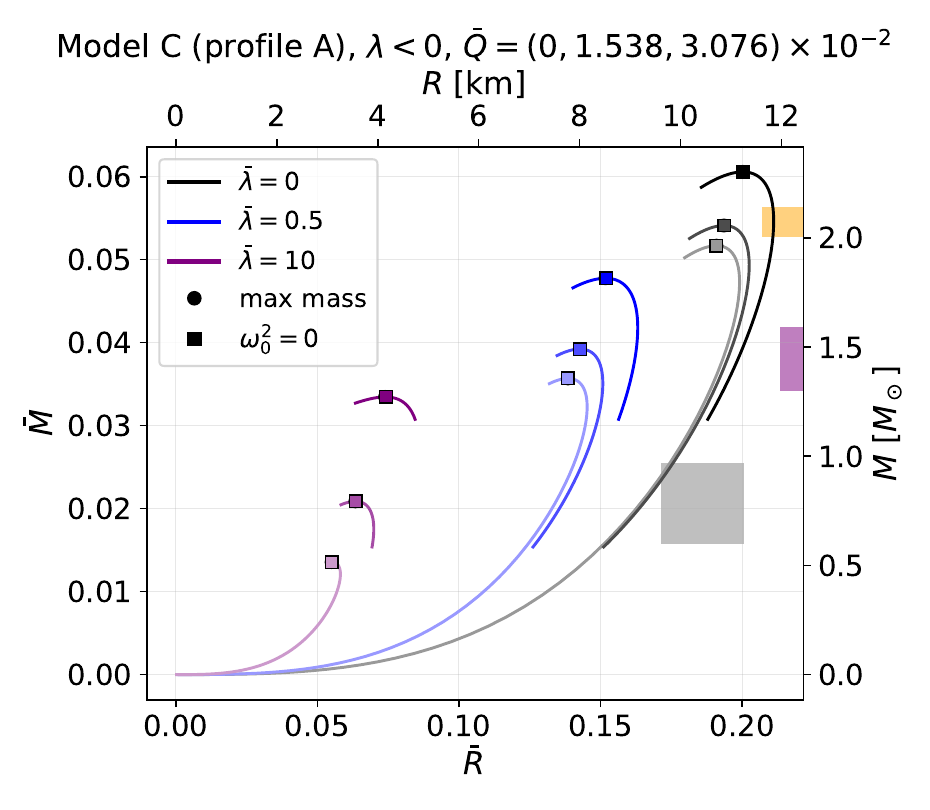}  
  \includegraphics[width=8 cm]{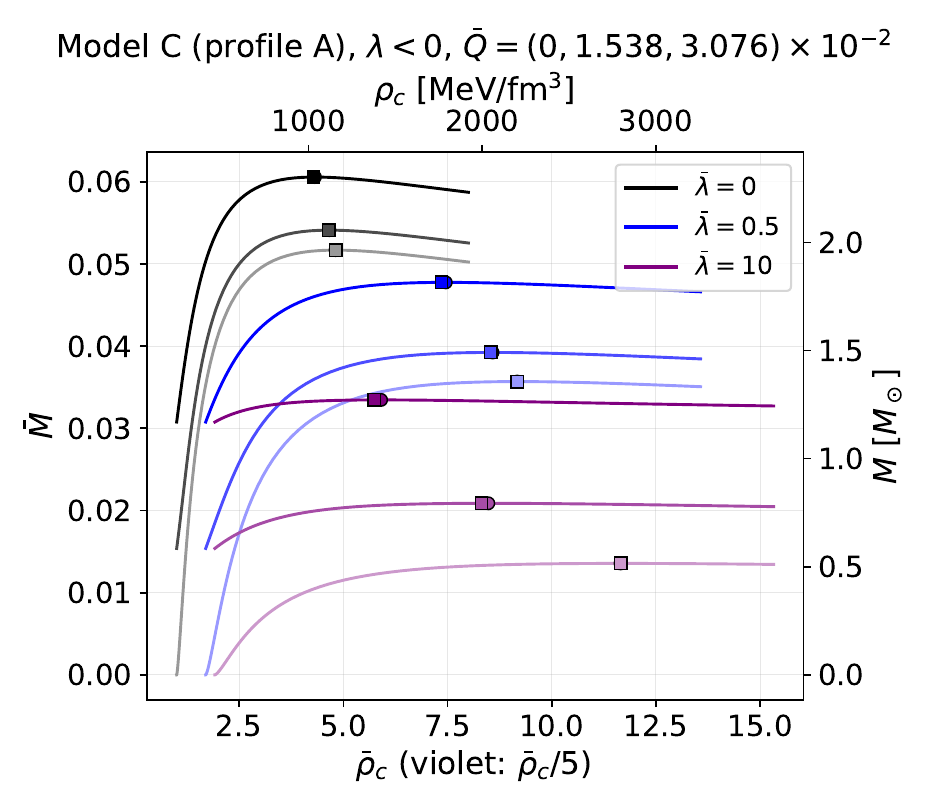}  
\caption{$\bar{M}$ vs. $\bar{R}$ (left) and $\bar{M}$ vs. $\bar{\rho}_c$ (right) of charged interacting quark stars in model C for $\bar{Q}=(0, 1.538,3.076)\times 10^{-2}$. The right and top axes in each plot are the corresponding dimensional results with $B_{\rm eff}=60\, \si{MeV/fm^3}$ for illustration and consistency with \cite{zhang:2021}. The  black, blue, and violet curves respectively denote  $\bar{\lambda}=(0, 0.5 , 10)$ with the sign of $\lambda$ being negative. Darker shades correspond to increasing values of $\bar{Q}$. The solid dots denote the maximum mass configurations, with filled squares denoting the point at which $\bar{\omega}^2_0=0$. Finally, the coloured boxes are $1\sigma$ observational estimates of mass/radii for PSR J0030+0451 (violet) \cite{Miller_2019}, PSR J0740+6620 (yellow) \cite{salmi2024}, HESS J1731-347 (gray) \cite{Horvath_2023}, and GW190814 (green) \cite{Abbott_2020}.  Note that for the violet curves in the $\bar{M}$ vs. $\bar{\rho}_c$ plots, we rescaled the x-axis as $\bar{\rho}_c\to\bar{\rho}_c/5$ for clarity.
}
\label{fig:mC_negLam}
\end{figure}

 When $\lambda<0$, the results for charge model C in the previous investigation \cite{zhang:2021} indicated that
separations between maximum mass and stability points are made larger as the magnitude of $\lambda$ increases, in concert with the charge, which also induces separation. Together this is manifest as  charge effects being most effective when $\lambda$ has a large magnitude and is negative. In our updated analysis we find that the behaviour is similar (see figure \ref{fig:mC_negLam}), but with the separation sizes generally being much smaller (again leading to more of the stars being stable than we previously thought). 

We have again included the four recent observational constraints (not present in \cite{zhang:2021}) in all of our $M$/$R$ diagrams for the $\lambda<0$ case as well. However in this case we find that  the observed mass-radius constraints for all four objects are not always consistent with the curves we obtain for the different charge models, due to negative $\lambda$ pushing results to smaller masses and radii).

\section{Discussion}\label{sec:discussion}

We have analyzed the
stability properties of charged quark stars using equation \eqref{SLform} with $\cal Q$ given by \eqref{Qcorr}, correcting the equation employed in a previous study  \cite{Goncalves:2020joq,zhang:2021}. The individual cases (consisting of charge model, charge strength, and size/sign of $\lambda$) in general differ in their effects on the separation between the stability point and maximum mass point, but some general trends can be observed.

First, the relative direction that the stability point moves with respect to the maximum mass point is unchanged from what was previously reported \cite{zhang:2021}. In model A (and C, which implicitly uses model A) the stability points move toward smaller central densities, whereas in model B they move to larger central densities. The size of the induced separation between the two points   in general differs from what was obtained previously in \cite{zhang:2021}. 

In general  the separation size in model A increases as the charge parameter $\alpha$ increases, but eventually turns around and closes for maximal $\alpha$. Making $\lambda$ more positive in these cases suppresses this separation, and accordingly making $\lambda$ more negative can contribute to the separation. Barring the turnaround at large charge which is unique to the present investigation, these general trends are in accordance with the earlier work \cite{zhang:2021} (however the separation sizes are smaller than previously reported in \cite{zhang:2021}).

For model B, separation size also increases as $\beta$ increases but we do not observe any kind of turnaround behaviour in the range of $\beta$ considered. The effects of $\lambda$ are slightly more complicated than the model A case -- making $\lambda$ more positive first contributes to an increased separation from $\bar{\lambda}=0\to\bar{\lambda}=0.5$, but then suppresses separation when increased further. On the other hand, making $\lambda$ more negative strictly suppresses the separation induced by $\beta$. It is worth noting that \cite{zhang:2021} also saw this turnaround behaviour for $\lambda>0$ under charge model B \cite{zhang:2021}, and negative $\lambda$ also had a suppressing effect there. The most notable difference is the fact that in this work all model B branches have stable stars, including when $\lambda\to\infty$ (which was not the case in \cite{zhang:2021}).

In general the separation sizes for the fixed charges considered in model C are very small in the present paper and are hardly perceptible on an $M$/$R$ plot, but since these solutions are implicitly using model A, we know that the qualitative effects of $\lambda$ and (small) charge parameter $\alpha$ in both our study and \cite{zhang:2021} match those already discussed for model A. The unique feature here (in contrast with \cite{zhang:2021}) is that once again all branches contain stable star solutions, even those starting relatively near the maximum mass point.

\section{Conclusion}\label{sec:conclusion}

We have numerically recalculated the stability properties of charged interacting quark stars, with the stability equations from \cite{Goncalves:2020joq} (propagating to \cite{zhang:2021}) corrected. Many of the general trends of the parameter space remain unchanged, but the actual magnitudes of the separations between the stability point and maximum mass can change significantly.

The most interesting qualitative changes to the results include ``turnaround" behaviour of the separation which reapproaches 0 for maximal model A charge parameter $\alpha$, and the fact that model B branches which were entirely unstable in \cite{zhang:2021} all contain stable star solutions in the corrected results. In all cases these changes allow for more stable stars than was reported in \cite{zhang:2021}.

 It is clear from \cref{fig:modelAresults,fig:modelBresults,fig:modelCresults,fig:mA_negLam,fig:mB_negLam,fig:mC_negLam} that, for particular choices of the strong interaction parameter (such as $\lambda > 0$ and $0\leq \bar{\lambda} \leq 1$), a subset of these solutions are stable and also pass through all four of the included observational constraints. A larger number of stable solutions satisfy only some of the constraints, and a larger amount still satisfy none. There is no guarantee that these objects actually are (charged) quark stars, however if they are, the data is consistent with an interacting equation of state for particular values of the strong interaction parameter $\lambda$.

\section*{Acknowledgements}
This work was supported in part by the Natural Sciences and Engineering Research Council of Canada.

\bibliographystyle{unsrt}

\cleardoublepage
\phantomsection  
\renewcommand*{\refname}{References}

\bibliography{bib}

\appendix
\section{Units and Scaling}\label{sec:appendix}
In all cases we numerically solve the TOV equations in a unitless form where the stellar variables have been rescaled using the effective bag constant $B_\mathrm{eff}$. More specifically, if we perform the following rescalings:
\begin{equation}
    \bar{\rho}=\frac{\rho}{4\,B_{\rm eff}}, \,\, \bar{P}=\frac{P}{4\,B_{\rm eff}}, \,\,\bar{\lambda}=\frac{\lambda^2}{4B_{\rm eff}}
\end{equation}
along with
\begin{align}
\bar{m}&=m{\sqrt{4\,B_{\rm eff}}}, \quad \bar{r}={r}{\sqrt{4\,B_{\rm eff}}},\\
\bar{q}&={q}{\sqrt{4\,B_{\rm eff}}}, \quad \bar{\rho}_e=\frac{\rho_e}{4\,B_{\rm eff}}
\end{align}
the TOV equations \eqref{eq:TOV1}-\eqref{eq:TOV4} can be rewritten in unitless form simply by replacing all variables with their unitless (barred) counterparts. Regarding the rescaling of charge model parameters, $\alpha$ is unitless to begin with. However $\beta$ and $Q$ rescale as
\begin{equation}
    \bar{\beta}=\frac{\beta}{4B_{\rm eff}}, \quad \bar{Q}={Q}{\sqrt{4\,B_{\rm eff}}}.
\end{equation}

\end{document}